\documentclass[twocolumn,showpacs,prl]{revtex4}
\usepackage{graphicx}
\usepackage{bm}
\usepackage{psfig}
\usepackage{amsmath}
\usepackage{amssymb}
\begin{document}

\title{Fractional Quantum Hall Effect at High Fillings in a Two-subband Electron System}
\author{J.~Shabani, Y.~Liu and M.~Shayegan}
\affiliation{Department of Electrical Engineering, Princeton University, Princeton, NJ 08544, USA}
\date{\today}
\begin{abstract}
Magneto-transport measurements in a clean two-dimensional electron system confined to a wide GaAs quantum well reveal that, when the electrons occupy two electric subbands, the sequences of fractional quantum Hall states observed at high fillings ($\nu > 2$) are distinctly different from those of a single-subband system. Notably, when the Fermi energy lies in the ground state Landau level of either of the subbands, no quantum Hall states are seen at the even-denominator $\nu$ = 5/2 and 7/2 fillings; instead the observed states are at $\nu = (i + p/(2p \pm 1))$ where $i$ = 2, 3, and $p$ = 1, 2, 3, and include several new states at $\nu$ = 13/5, 17/5, 18/5, and 25/7.

\end{abstract}


\maketitle

The ground states of low-disorder two-dimensional electron systems (2DESs) at high Landau level (LL) fillings ($\nu >
2$) have been enigmatic. Early experiments provided evidence for a unique fractional quantum Hall state (FQHS) at the
even-denominator filling $\nu$ = 5/2 \cite{Willett87}. More recent measurements on the highest quality 2DESs have
revealed a plethora of additional ground states including insulating and density-modulated phases
\cite{Lilly.PRL.1999,Pan.PRL.1999,Eisenstein.PRL.2002,Xia.PRL.2004,Choi.PRB.2008,Pan.PRB.2008,Dean.PRL.Oct08}. But
absent are clear sequences of odd-denominator FQHSs at $\nu = i + p/(2p\pm1)$ (where $p$ = 1, 2, 3, ...) that are
typically observed at lower fillings (i.e., when $i$ = 0 or 1) \cite{Jain.CF.book}. It is believed that, in the higher
LLs, the larger extent of the electron wavefunction (in the 2D plane), combined with the presence of extra nodes, leads
to a modification of the (exchange-correlation) interaction effects and stabilizes the non-FQHSs at the expense of
FQHSs.

Meanwhile, the origin and the stability of the FQHSs at high fillings, especially those at $\nu$ = 5/2 and 12/5, have
become the focus of renewed interest since these states might obey non-Abelian statistics and be useful for topological
quantum computing \cite{Nayak08}. In particular, it has been proposed that the $\nu$ =5/2 FQHS should be particularly
stable in a "thick" 2DES confined to a relatively wide quantum well (QW) \cite{Peterson.PRB.08}. In a realistic,
experimentally achievable system, of course, the electrons in a wide QW typically occupy two (or more) electric subbands
\cite{Peterson.PRB.10}. Here we report measurements in such a system. Figure 1 highlights our main
observations. In contrast to data taken in a narrow (30 nm) GaAs QW where only one electric subband is occupied (Fig.
1(a)), data for the wider (56 nm) well (Figs. 1(b,c)) \cite{Footnote1} do not exhibit even-denominator states at $\nu$ =
5/2 and 7/2. Instead, we observe FQHS sequences at $\nu = 2 + p/(2p \pm 1)$ and $3 + p/(2p \pm 1)$,
reminiscent of the usual composite Fermion (CF) sequences observed at lower $\nu$ around 1/2 and 3/2 (i.e.,
at $\nu = 0 + p/(2p \pm 1)$ and $\nu = 1 + p/(2p \pm 1)$) \cite{Jain.CF.book}. The FQHSs we observe include states at
$\nu$ = 7/3, 8/3, 12/5, 13/5, 10/3, 11/3, 17/5, 18/5, and 25/7, some of which have not been previously seen
\cite{Pan.PRB.2008}.

Our samples were grown by molecular beam epitaxy and consist of GaAs QWs bounded on each side by undoped
Al$_{0.24}$Ga$_{0.76}$As spacer layers and Si $\delta$-doped layers. We studied several samples with well widths ($w$)
ranging from 30 to 80 nm. Here we focus on data from two samples; a narrow QW ($w = $ 30 nm) in which the electrons
occupy one electric subband, and a wide ($w = $ 56 nm) QW where two subbands are occupied. The
low-temperature mobility in our single-subband samples is in excess of $\simeq$ 1000 m$^{2}$/Vs, while the two-subband
samples have mobilities which are typically about two to three times smaller. Since our samples were grown under very
similar conditions it appears that the lower mobility in the wider samples is a consequence of the occupancy of the
second subband. We used an evaporated Ti/Au front-gate and an In back-gate to change the 2DES density $n$ and tune the
charge distribution symmetry. The transport traces reported here were all measured in a dilution refrigerator at a
temperature of $\simeq$ 30 mK.

In wide QW samples, the electrons typically occupy two electric subbands, separated in energy by an amount which we
denote $\Delta$. When the QW is "balanced," i.e., the charge distribution is symmetric, the occupied subbands are the
symmetric (S) and anti-symmetric (AS) states. When the QW is "imbalanced," the two occupied subbands are no longer
symmetric or anti-symmetric; nevertheless, for brevity, we still refer to these as S (ground state) and AS (the excited
state). In our experiments, we carefully control the electron density ($n$) and charge distribution symmetry in the wide
QW via applying back and front gate biases \cite{Suen94,Shabani.PRL.Dec.09}. For each pair of gate biases, we measure
the occupied subband electron densities from the Fourier transforms of the low-field ($B \leq$ 0.4 T) magneto-resistance
oscillations. These Fourier transforms exhibit two peaks whose frequencies are directly proportional to the densities of
the two occupied subbands (see, e.g., Fig. 1 in Ref. \cite{Shabani.PRL.Dec.09}). The difference between these
frequencies is therefore a direct measure of $\Delta$. Note that, at a fixed $n$, $\Delta$ is smallest when the charge
distribution is balanced and it increases as the QW is imbalanced. By monitoring the evolution of these frequencies as a
function of $n$ and, at a fixed $n$, as a function of the back- and front-gate biases, we can tune the symmetry of the
charge distribution \cite{Suen94,Shabani.PRL.Dec.09} and also precisely determine the value of $\Delta$. Throughout this
Letter, we quote the experimentally measured values of $\Delta$. Another experimentally determined relevant parameter is
the charge imbalance $\delta n$, defined as the amount of charge transferred from the back side of the QW to the front
side. We note that our measured $\Delta$ for given values of $n$ and $\delta n$ are in very good agreement with the
results of our self-consistent calculations of charge distribution and energy levels in our wide QW. We show examples of
such calculations for a balanced and an imbalanced charge distribution in Fig. 2(a) insets.

\begin{figure}[ht!]
\centering
\includegraphics[scale=0.9]{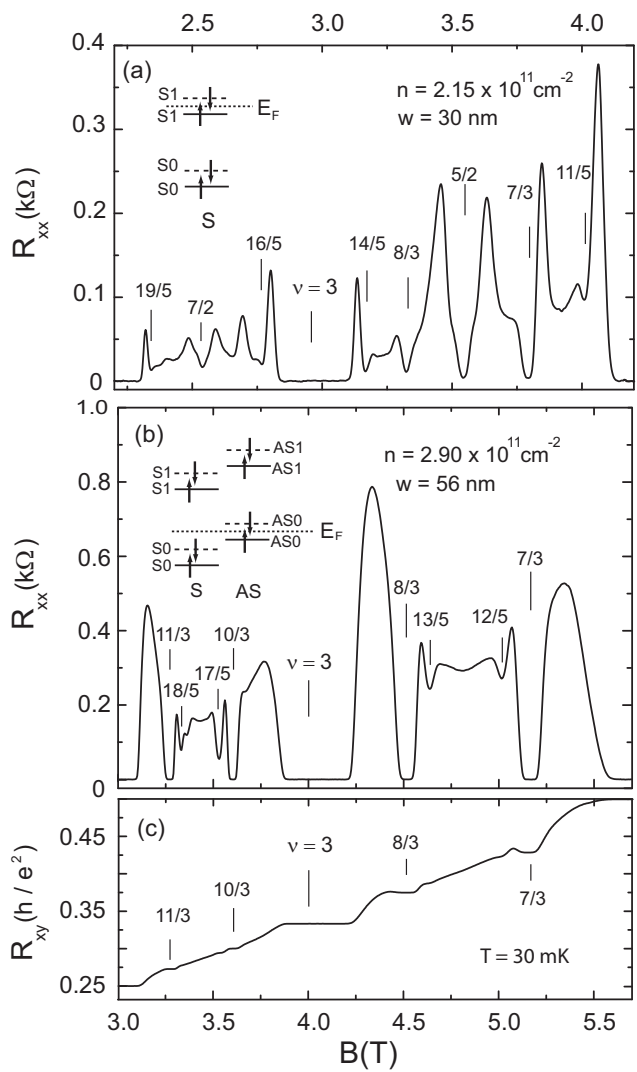}
\caption{Longitudinal magneto-resistance ($R_{xx}$) traces, showing FQHSs in the range $2 < \nu < 4$ for: (a) a 30
nm-wide, and (b) a 56 nm-wide QW sample. The insets schematically show the positions of the spin-split LLs of the lowest
(S) and the second (AS) electric subbands, as well as the position of the Fermi energy ($E_F$) at $\nu = 3$; the indices
0 and 1 indicate the lowest and the excited LLs, respectively. The vertical lines mark the expected field positions of
various fractional fillings. (c) Hall resistance corresponding to the data of (b). }

\end{figure}

\begin{figure*}[ht!]
\centering

\includegraphics[scale=1.1]{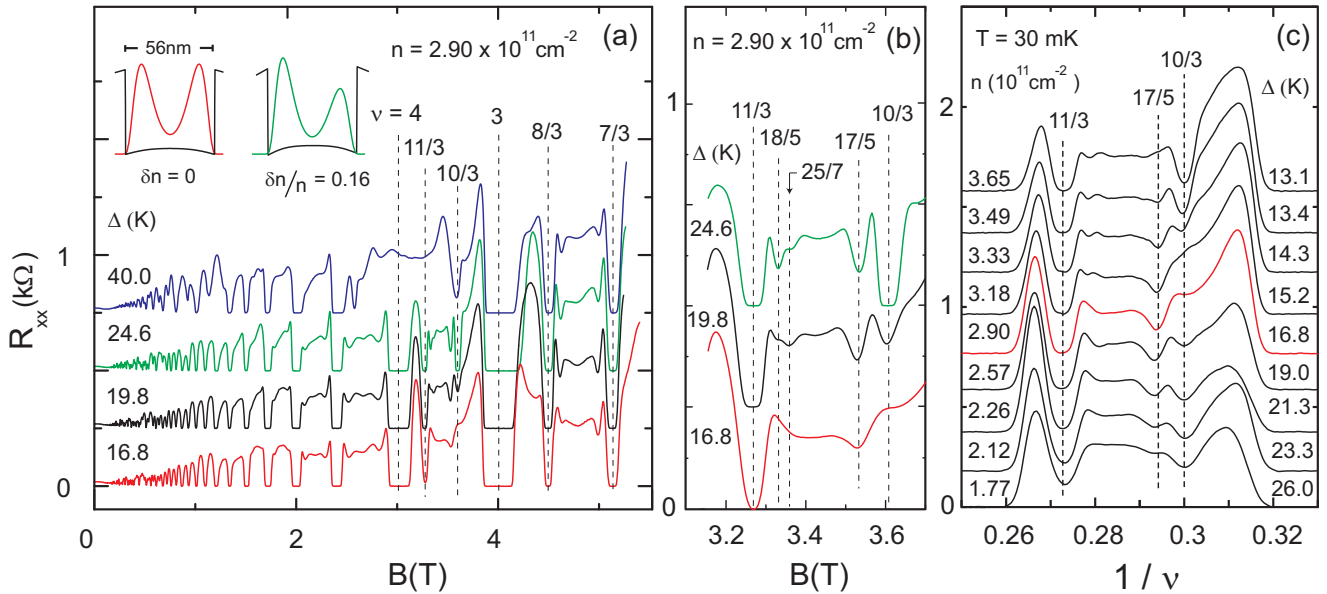}
\caption{(Color online) (a) and (b) Evolution of $R_{xx}$ vs $B$ traces and the FQHSs for the 56 nm-wide QW sample at a fixed
density $n = 2.90 \times 10^{11}$ cm$^{-2}$. The bottom (red) trace is for the balanced case ($\delta n = 0$), and the
other traces are for increasingly imbalanced charge distributions. The measured subband separation ($\Delta$) for each
trace is indicated on the left. Insets: Calculated charge distribution and potential (at zero magnetic field) at $n =
2.90 \times 10^{11}$ cm$^{-2}$ for $\delta n/n = 0$ and $\delta n/n = 0.16$. (c) Evolution of FQHSs in the range
$3 < \nu < 4$ with density. For each trace the charge distribution is kept symmetric; the densities are indicated on the left and the measured values of $\Delta$ on the right.}
\end{figure*}

Figure 2(a) captures the evolution of $R_{xx}$ traces taken for the 56 nm-wide QW sample as the charge distribution is
imbalanced and $\Delta$ is increased. All the traces were taken at a fixed $n = 2.90 \times 10^{11}$ cm$^{-2}$ while the
charge distribution was made increasingly more asymmetric so that more charge resided near the front interface of the
QW. Traces taken for the opposite direction of imbalance, i.e., when the electrons were pushed toward the back
interface, show a very similar behavior. We emphasize that the 2DES mobility decreases by less than 10\% as the charge
is imbalanced so changes in disorder cannot explain the evolution of FQHSs seen in Fig. 2.

The most striking features of Fig. 2(a) data are the sequences of FQHSs observed at $\nu = i + p/(2p \pm 1)$ for $i = 2$
and 3. Also remarkable is the evolution of these FQHSs as a function of increasing the charge imbalance (and therefore
$\Delta$) at fixed $n$ (Figs. 2(a,b) and 3), or changing the total density (Fig. 2(c)), as we discuss later in the paper.
Note also the absence of FQHSs at $\nu$ = 5/2 and 7/2, typically seen in very high mobility 2DESs confined to narrower
GaAs QWs (e.g., see Fig. 1(a)).

To discuss these observations, we consider two other relevant energies, the cyclotron energy ($E_{C} = \hbar eB/m^{*}$)
and the Zeeman energy ($E_{Z} = \mu_{B} |g^{*}| B$), where $m^{*}$ and $g^{*}$ are the effective mass and Lande
g-factor. Assuming the GaAs band values ($m^{*} = 0.067 m_{0}$ and $g^{*} = -0.44$), we have $E_{C} = 20 \times B$ and
$E_{Z} = 0.30 \times B$ in units of K, where $B$ is in T. In a typical (narrow) GaAs QW, $\Delta > E_{C} > E_{Z}$, so
that for $2 < \nu < 4$ the Fermi energy ($E_F$) lies in the excited orbital LL of the lowest electric subband (i.e., S1;
see Fig. 1(a) inset). In our wide QW, however, $\Delta$ and $E_{Z}$ are both smaller than $E_{C}$ in the range $2 < \nu
< 4$, so that a situation like the one shown in Fig. 1(b) ensues, where $E_F$ lies in the $lowest$ (orbital) LLs of the
two electric subbands (i.e., S0 and AS0). It is not a priori obvious whether $\Delta$ is smaller or larger than $E_{Z}$
in our sample for $2 < \nu < 4$ since both $\Delta $ and $E_Z$ can be re-normalized because of interaction. If we use
the band value of the g-factor, $E_{Z} < \Delta$. However, at $n = 2.90 \times 10^{11}$ cm$^{-2}$, we observe a
disappearance of the integer quantum Hall state at $\nu = 4$ when $\Delta \simeq$ 40 K (top trace in Fig. 2(a)).
Associating this disappearance with the coincidence of the AS0$\downarrow$ and S1$\uparrow$ LLs (Fig. 1(b)), expected
when $E_{C} = E_{Z} + \Delta$, we find $E_{Z} + \Delta = $ 60 K at the field position of $\nu$ = 4 ($B \simeq 3$ T),
implying that $E_{Z}$ and/or $\Delta$ are enhanced compared to their low field values; such enhancements have been
previously reported for 2DESs in wide GaAs QWs \cite{Muraki.PRL.01,Solovyev.PRB.09}.

Regardless of the relative magnitudes of $E_{Z}$ and $\Delta$, it is clear that for the bottom three traces of Fig.
2(a), for $2 < \nu < 4$, $E_F$ lies in the {\it lowest orbital} LLs of the two electric subbands (i.e., S0 and AS0,
see Fig. 1(b)). We believe this is the reason why our wide QW sample does not exhibit even-denominator FQHSs at $\nu =
5/2$ and 7/2, and instead shows FQHS sequences that are typically seen in a narrow QW at lower fillings ($\nu < 2$) when
$E_F$ also lies in the lowest orbital LL \cite{Jain.CF.book}. Strong evidence for this conjecture is provided in
Fig. 3, where we present data at a higher density $n = 3.65 \times 10^{11}$ cm$^{-2}$ and as the QW is made extremely
imbalanced. At low and moderate imbalances ($\Delta < $ 30 K), for $3 < \nu < 4$, $E_F$ lies in the AS0$\downarrow$ level (see Fig. 3(d)), and the data are qualitatively similar to those in Fig. 2(b), i.e., the $\nu = (3 + p/(2p \pm 1))$ FQHSs are observed. For $\Delta \simeq$ 42 K, the $\nu =
4$ $R_{xx}$ minimum completely disappears, signaling a coincidence of the S1$\uparrow$ and AS0$\downarrow$ LLs at this
filling (see Fig. 3(c)). As we further imbalance the QW past the coincidence (top two traces in Fig. 3(a)), $E_F$ for $3 < \nu < 4$ lies
in the $excited$ LL of the symmetric subband (i.e., S1$\uparrow$, see Fig. 3(b)). Consistent with our conjecture, in this case there is
no strong FQHS sequence at $\nu = (3 + p/(2p \pm 1))$ and instead the even-denominator $\nu = 7/2$ state is observed,
similar to the data of Fig. 1(a) for a narrow QW.

It is worth emphasizing that the FQHSs we observe cannot be viewed as simple combinations of two FQHSs (each at $\nu
/2$) in two parallel layers; this is obvious for the odd-numerator states at $\nu = 7/3$ and 11/3, as there are no FQHSs
at 7/6 or 11/6 fillings \cite{FootnoteA}. We add that, as qualitatively clear from the data of Figs.~1-3, the energy gaps for the
odd-numerator FQHSs we observe are typically much larger when these states are formed in the upper LLs. For example,
from the temperature dependence of the $R_{xx}$ minima at $\nu =$ 10/3 and 11/3 states in $\Delta = $ 24.6 K trace of Fig.~2(b), we measure a gap of $\simeq$ 0.8 K, clearly much larger than in the top trace where these states are barely
developed \cite{FootnoteB}.

Data of Figs. 2 and 3 further demonstrate that, even before the $\nu$ = 4 electron LL coincidence occurs, the $\nu = i + p/(2p
\pm 1)$ FQHSs exhibit a subtle evolution as $\Delta$ is increased. In the $3 < \nu < 4$ range, e.g., the $\nu = 11/3$
FQHS is always present and relatively strong (except at and past the $\nu$ = 4 coincidence), while the strengths of the 10/3 state, as well as
the weaker 17/5, 18/5 and 25/7 states, critically depend on $\Delta$. A qualitatively similar evolution is also observed
when the charge distribution is kept symmetric but $n$ is varied (Fig.~2(c)). These evolutions resemble those seen for
the FQHSs in the $1 < \nu < 2$ range in a narrow GaAs QW as a function of spin polarization
\cite{Jain.CF.book,Du.PRL.75.3926.95,Park.PRL.98}, or in an AlAs QW as a function of valley polarization
\cite{Padmanabhan.PRB.80.035423.09}. In those cases, the observations can be explained in terms of LLs for
two-component CFs which have either a spin or valley degree of freedom, and the coincidences of these LLs as the degree
of CFs' spin or valley polarization is tuned.

We believe the evolutions seen in Figs. 2 and 3 likely have a similar origin. One possibility is to interpret the data in terms of two-component CFs which are formed in the AS0 LL and have a spin degree of freedom. In such a picture, the lowest (S0) LLs are completely filled and inert, so that the FQHSs in the filling range $3 < \nu < 4$ correspond to integer quantum Hall states of CFs with filling $p$, with the expression $\nu = 2 + (2 - p/(2p \pm 1))$ giving the relation between $\nu$ and $p$. This expression, in which the first term 2 accounts for the two S0 LLs being inert and the second term 2 takes into account particle-hole symmetry, maps the $\nu = 11/3$ FQHS to $p = 1$, the $\nu = 10/3$ and 18/5 states to $p = 2$, and the $\nu = 17/5$ and 25/7 states to $p = 3$. The evolution of the FQHSs can then be explained as coincidences of the CF LLs, similar to what has been reported for electrons in the S0 LLs in single-subband 2D systems \cite{Du.PRL.75.3926.95,Park.PRL.98,FootnoteC}. Another possibility is that the evolution we observe stems from an interplay between the CFs' spin $and$ subband degrees of freedom which leads to four-component CFs. In this scenario, all the four lowest levels (S0$\uparrow$, S0$\downarrow$, AS0$\uparrow$, and AS0$\downarrow$) would be relevant for the formation of the CFs, and the mapping of the FQHS fillings in the range $3 < \nu < 4$ and the corresponding CF integer fillings $p$ is given through the expression $\nu = (4 - p/(2p \pm 1))$. This expression, which takes into account the particle-hole symmetry in a four-component CF system, maps the FQHSs in the range $3 < \nu < 4$ to the same $p$ as in the above two-component picture. We note that our data are qualitatively consistent with either of these CF LL pictures; we defer a detailed comparison of the data with the predictions of these models to a future communication.

\begin{figure}[ht!]
\centering
\includegraphics[scale=1]{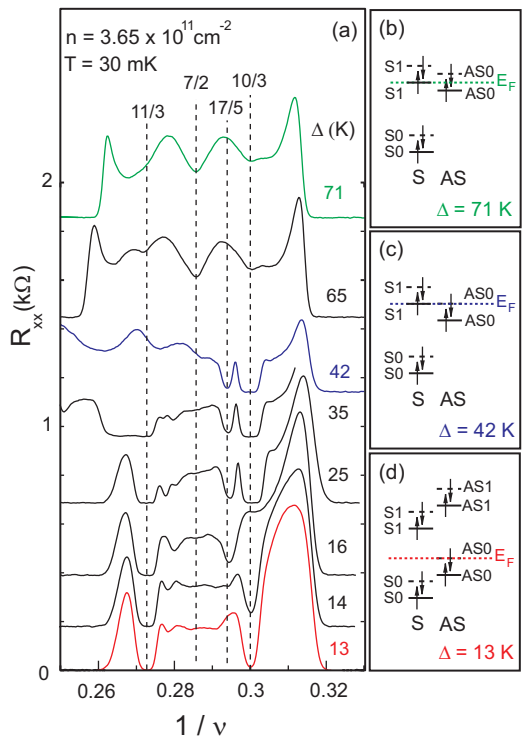}
\caption{(a) (Color online) (a) Evolution of the FQHSs in the range $3 < \nu < 4$ at a fixed density of $n = 3.65 \times 10^{11}$ cm$^{-2}$ with charge imbalance. The bottom (red) trace is for the balanced case, and the other traces are for increasingly imbalanced charge distributions. The measured $\Delta$ for each trace is indicated on the right. (b-d) Landau level diagrams schematically showing the position of $E_F$ for three values of $\Delta$ as indicated. }
\end{figure}

The results presented here demonstrate that the stability of the FQHSs in the filling range $2 < \nu < 4$ crucially depends on whether $E_F$ resides in the lowest or the excited orbital LLs. When $E_F$ lies in an excited LL, as is the case in the standard (narrow) QWs, the even-denominator states are stable. But if two electric subbands are occupied so that $E_F$ resides in the ground state LLs of these subbands, then the even-denominator states are absent and instead FQHSs are seen at the CF filling sequence $\nu = (i + p/(2p \pm 1))$. Our data also reveal a subtle evolution of the FQHSs in the range $2 < \nu < 4$ with changes in density and/or subband separation, suggesting coincidences of CF LLs.

We thank J.K. Jain and C. Toke for illuminating discussion. We acknowledge support through the NSF (DMR-0904117 and MRSEC DMR-0819860) for sample fabrication and characterization, and the DOE BES (DE-FG02-00-ER45841) for measurements.


\begin{thebibliography}{99}

\small

\bibitem{Willett87}
R.L. Willett {\it et al.}, Phys. Rev. Lett. {\bf 59}, 1776 (1987).

\bibitem{Lilly.PRL.1999}
M. P. Lilly {\it et al.}, Phys. Rev. Lett. {\bf 82}, 394 (1999).

\bibitem{Pan.PRL.1999}
W. Pan {\it et al.}, Phys. Rev. Lett. {\bf 83}, 3530 (1999).

\bibitem{Eisenstein.PRL.2002}
J. P. Eisenstein {\it et al.}, Phys. Rev. Lett. {\bf 88}, 076801 (2002).

\bibitem{Xia.PRL.2004}
J. S. Xia {\it et al.}, Phys. Rev. Lett. {\bf 93}, 176809 (2004).

\bibitem{Pan.PRB.2008}
W. Pan {\it et al.}, Phys. Rev. B {\bf 77}, 075307 (2008).

\bibitem{Choi.PRB.2008}
H. C. Choi {\it et al.}, Phys. Rev. B {\bf 77}, 081301(R) (2008).

\bibitem{Dean.PRL.Oct08}
C. R. Dean {\it et al.},  Phys. Rev. Lett. {\bf 101}, 186806 (2008).

\bibitem{Jain.CF.book}
J. K. Jain, {\it Composite Fermions}, (Cambridge University Press, New York, 2007).

\bibitem{Nayak08}
C. Nayak {\it et al.}, Rev. Mod. Phys. {\bf 80}, 1083 (2008).

\bibitem{Peterson.PRB.08}
M. R. Peterson, Th. Jolicoeur and S. Das Sarma, Phys. Rev. B {\bf 78}, 155308 (2008).

\bibitem{Peterson.PRB.10}
The stability of the 5/2 state in such QWs has also been theoretically discussed [M.R. Peterson and S. Das Sarma, Phys.
Rev. B {\bf 81}, 165304 (2010)]. However, the relevance of this work to ours is unclear since it assumes a fully
spin-polarized 2DES.

\bibitem{Footnote1}
Figure 1(b) trace is for an imbalanced charge distribution with a subband separation of 24.6 K (see Fig. 2).

\bibitem{Suen94}
Y.W. Suen {\it et al.}, Phys. Rev. Lett. {\bf 72}, 3405 (1994).

\bibitem{Shabani.PRL.Dec.09}
J. Shabani {\it et al.}, Phys. Rev. Lett. {\bf 103}, 256802 (2009).

\bibitem{Muraki.PRL.01}
K. Muraki {\it et al.}, Phys. Rev. Lett. {\bf 87}, 196801 (2001).

\bibitem{Solovyev.PRB.09}
V.V. Solovyev {\it et al.}, Phys. Rev. B {\bf 80}, 241310 (2009).

\bibitem{FootnoteA}
A close examination of the evolution of the FQHSs we observe rules out that their origin is a magnetic field induced electron redistribution in the QW as discussed, e.g., in Ref. \cite{Solovyev.PRB.09}.

\bibitem{FootnoteB}
Similarly, we measure a gap of $\simeq$ 1.2 K for the 7/3 state in Fig. 1(b) compared to only $\simeq$
0.1 K in Fig. 1(a).

\bibitem{Du.PRL.75.3926.95}
R. R. Du {\it et al.}, Phys. Rev. Lett. {\bf 75}, 3926 (1995).

\bibitem{Park.PRL.98}
K. Park and J.K. Jain, Phys. Rev. Lett. {\bf 80}, 4237 (1998).

\bibitem{Padmanabhan.PRB.80.035423.09}
M. Padmanabhan {\it et al.}, Phys. Rev. B {\bf 80}, 035423 (2009).

\bibitem{FootnoteC}
Note that changes in the wavefunction shape can modify the Coulomb energy and hence tune the spin polarization of the CFs [C. Toke and J.K. Jain, unpublished; also, see S. Kraus {\it et al.}, Phys. Rev. Lett. {\bf 89}, 266801 (2002).]



\end{thebibliography}
\end{document}